\documentclass[aps,prl,preprint,bibliography]{revtex4-1}
\pdfoutput=1

\usepackage{gensymb}
\usepackage{color}
\usepackage{graphicx}
\usepackage{epstopdf}
\usepackage{dcolumn}
\usepackage{bm}
\usepackage[normalem]{ulem}
\usepackage{amsmath}
\usepackage{amssymb}
\usepackage{here}
\usepackage{color}
\usepackage{placeins}
\usepackage{ulem}
\usepackage{xcolor}
\DeclareRobustCommand{\erase}{\bgroup\markoverwith{\textcolor{red}{\rule[.5ex]{2pt}{0.4pt}}}\ULon}

\newcommand{\be}{\begin{equation}}
\newcommand{\ee}{\end{equation}}
\newcommand{\bfig}{\begin{figure}}
\newcommand{\efig}{\end{figure}}

\newcommand{\RRS}{\textit{R}Ru$_3$Si$_2$}
\newcommand{\RRB}{\textit{R}Ru$_3$B$_2$}
\newcommand{\NRB}{NdRu$_3$B$_2$}
\newcommand{\YRS}{YRu$_3$Si$_2$}
\newcommand{\LRS}{LaRu$_3$Si$_2$}

\newcommand{\PRS}{PrRu$_3$Si$_2$}
\newcommand{\NRS}{NdRu$_3$Si$_2$}

\newcommand{\AVS}{\textit{A}V$_3$Sb$_5$}
\newcommand{\KVS}{KV$_3$Sb$_5$}
\newcommand{\FG}{FeGe}
\newcommand{\SVS}{ScV$_6$Sb$_6$}
\newcolumntype{C}[1]{>{\centering\arraybackslash}p{#1}}

\begin{document}
\title{Successive orthorhombic distortions in kagome metals\\
by molecular orbital formation}

\author{Ryo Misawa$^{1}$}\email{misawann6@g.ecc.u-tokyo.ac.jp}
\author{Shunsuke Kitou$^{2}$}
\author{Rinsuke Yamada$^{1}$}
\author{Tobi Gaggl$^{1}$}
\author{Ryota Nakano$^{1}$}
\author{Yudai Shibata$^{1}$}
\author{Yoshihiro Okamura$^{1}$}
\author{Markus Kriener$^{3}$}
\author{Yuiga Nakamura$^{4}$}
\author{Yoshichika \=Onuki$^{3}$}
\author{Youtarou Takahashi$^{1}$}
\author{Taka-hisa Arima$^{2,3}$}
\author{Milena Jovanovic$^{5}$}
\author{Leslie M. Schoop$^{6}$}
\author{Max Hirschberger$^{1,3}$}\email{hirschberger@ap.t.u-tokyo.ac.jp}

\affiliation{$^{1}$Department of Applied Physics and Quantum-Phase Electronics Center (QPEC), The University of Tokyo, Bunkyo-ku, Tokyo 113-8656, Japan}
\affiliation{$^{2}$Department of Advanced Materials Science, The University of Tokyo, Kashiwa, Chiba 277-8561, Japan}
\affiliation{$^{3}$RIKEN Center for Emergent Matter Science (CEMS), Wako, Saitama 351-0198, Japan}
\affiliation{$^{4}$Japan Synchrotron Radiation Research Institute (JASRI), SPring-8, Hyogo 679-5198, Japan}
\affiliation{$^{5}$Department of Chemistry, North Carolina State University, Raleigh, NC 27695-8204, USA}
\affiliation{$^{6}$Department of Chemistry, Princeton University, Princeton, New Jersey 08544, USA}

\begin{abstract}
The kagome lattice, with its inherent frustration, hosts a plethora of exotic phenomena, including the emergence of $3\bm{q}$ charge density wave order.
The high rotational symmetry, required to realize such an unconventional charge order, is broken in many kagome materials by orthorhombic distortions at high temperature, the origin of which is much less discussed despite their ubiquity.
In this study, synchrotron X-ray diffraction reveals a structural phase transition from a parent hexagonal phase to an orthorhombic ground state, mediated by a critical regime of diffuse scattering in the prototypical kagome metals \RRS{} (\textit{R}=rare-earth).
Structural analysis uncovers an interlayer dimerization of kagome atoms in the low-temperature phase. Accordingly, a dimer model with one-dimensional disorder on kagome layers successfully reproduces the diffuse scattering. The observations point to molecular orbital formation between kagome $4d_{z^2}$ orbitals as the driving force behind the transition,
consistent with \textit{ab initio} calculations.
A framework based on electronegativity and atomic radii is proposed to evaluate the stability of the hexagonal phase in kagome metals, guiding the design of highly symmetric materials.

\end{abstract}

\date{\today}

\maketitle

\begin{figure*}[t]
  \begin{center}
		\includegraphics[clip, trim=0cm 0cm 0cm 0cm, width=0.7\linewidth]{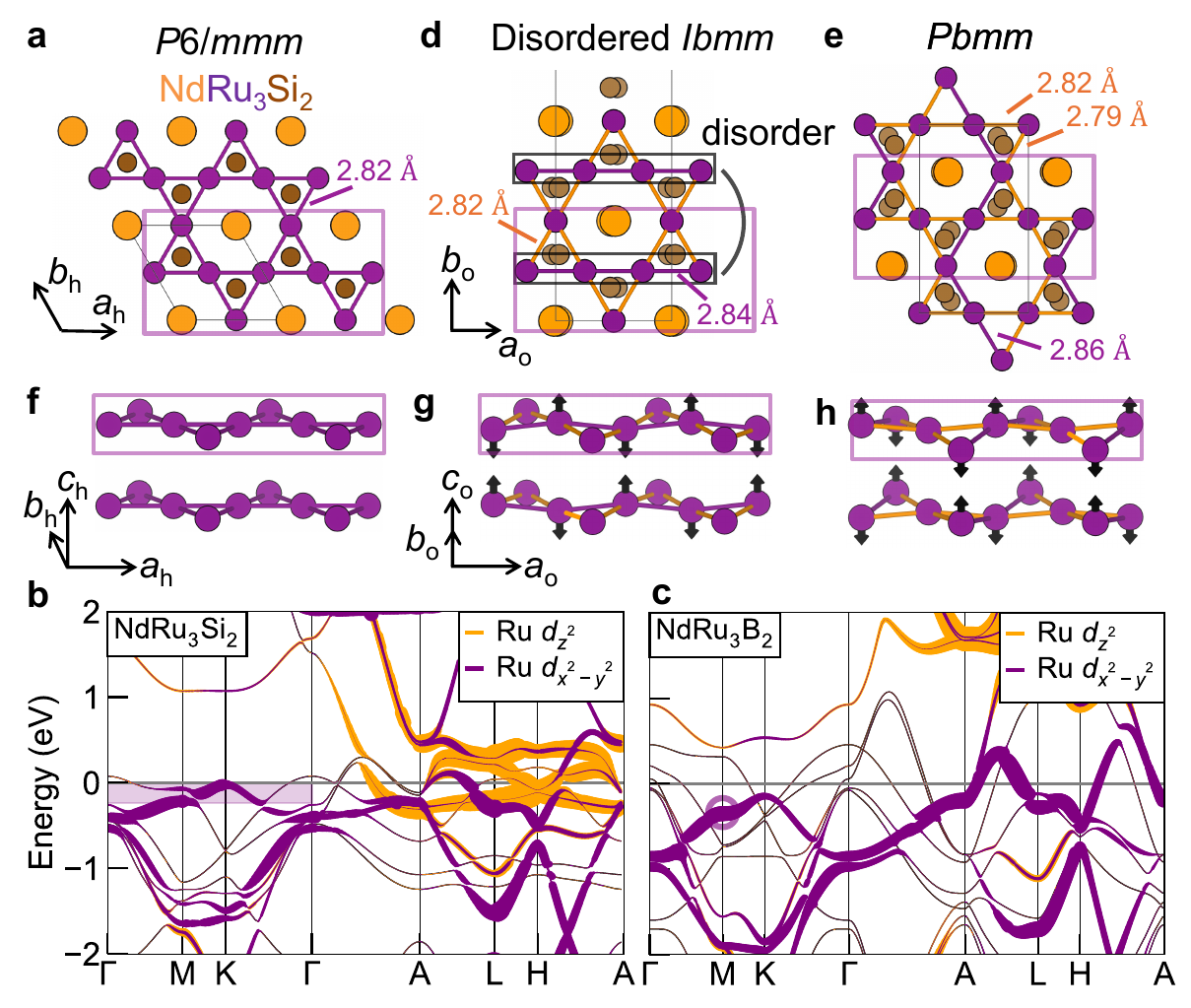}
    \caption[]{(color online). Successive orthorhombic distortions in kagome metals \RRS{} (\textit{R}=rare-earth).
    \textbf{a}, Hexagonal structure in space group $P6/mmm$ above $T_\mathrm{str1} = 760\,$K in \NRS{}. Ru atoms form the kagome lattice. \textbf{b}, Band structure of \NRS{} with flat bands near the Fermi level in the hexagonal phase as highlighted by the purple area.  Yellow and purple colors represent $d_{z^2}$ and $d_{x^2-y^2}$ orbitals, respectively.
    \textbf{c}, Band structure of \NRB{} with van Hove singularity near the Fermi energy, highlighted by the purple ring.
    \textbf{d}, Disordered $Ibmm$ structure of \NRS{} between $T_\mathrm{str1}$ and $T_\mathrm{str2} = 720\,$K. Ru atoms dimerize along the $c_\mathrm{o}$ axis. The dimers are periodically arranged along the $a_\mathrm{o}$ axis but randomly along the $b_\mathrm{o}$ axis.
    Blue and red colors highlight short and long Ru-bonds, respectively.
    \textbf{e}, Orthorhombic structure with $Pbmm$ below $T_\mathrm{str2} = 660\,$K. The disordered structure of Ru-Ru dimers, similar to panel \textbf{d}, remains in an intermediate temperature range. Compared to the average $Ibmm$ structure in panel \textbf{d}, the Ru-Ru dimers are recombined, as seen from the zigzag pattern of the purple bonds.
    \textbf{f}-\textbf{h}, Side views of $P6/mmm$, $Ibmm$, and $Pbmm$ structures. Arrows indicate the direction of displacement. Subscripts h and o of the unit cell axes denote hexagonal and orthorhombic phases, respectively. Purple regions correspond to those in panels \textbf{a},\textbf{d}, and \textbf{e}.
     }
    \label{fig:fig1}
  \end{center}
\end{figure*}
\begin{figure*}[t]
  \begin{center}
		\includegraphics[clip, trim=0cm 0cm 0cm 0cm, width=1.\linewidth]{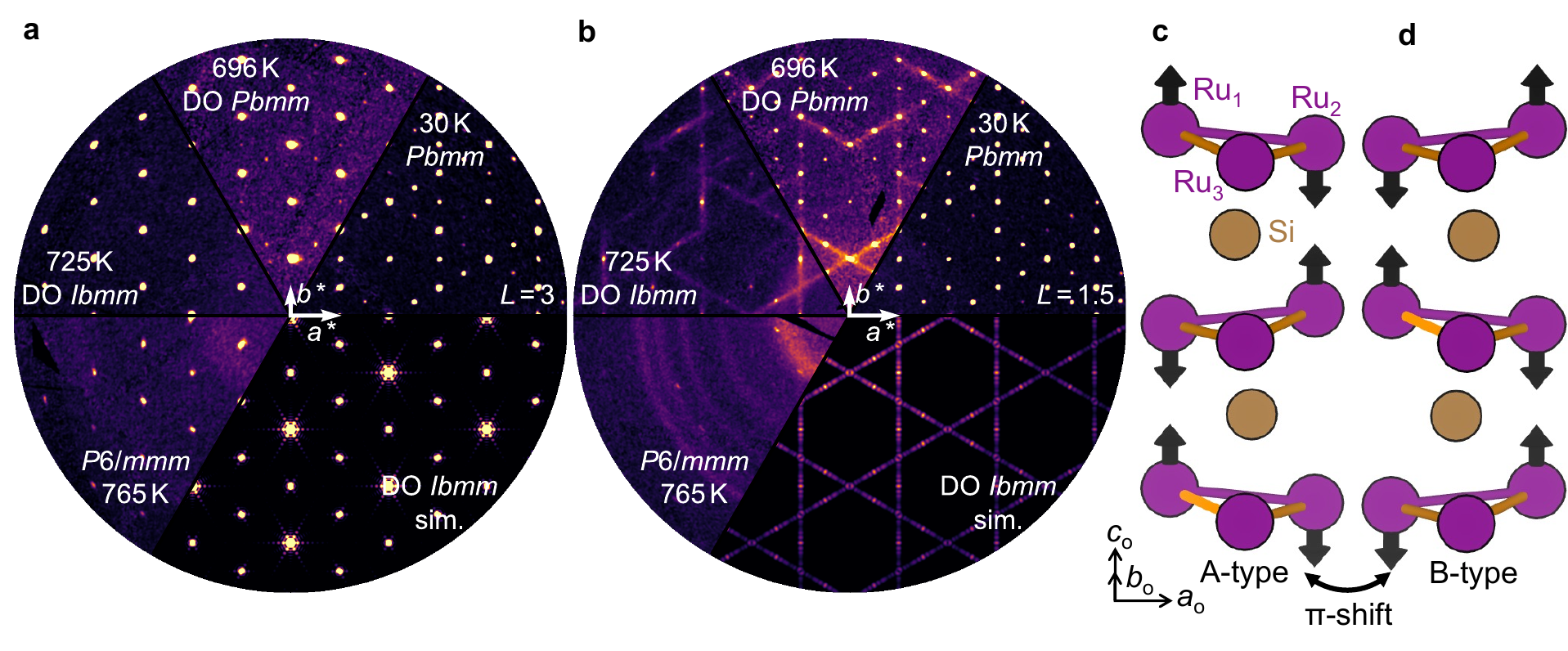}
    \caption[]{(color online). Temperature-dependent reciprocal space map and modeling of the disordered structure in \NRS{}. \textbf{a}, XRD pattern at the $(H\,K\,3)$ plane in the orthorhombic basis from four distinct phases: $P6/mmm$ ($765\,$K), disordered (DO) $Ibmm$ ($725\,$K), disordered $Pbmm$ ($696\,$K), and $Pbmm$ ($30\,$K). The simulated pattern is for the disordered $Ibmm$ phase.
    \textbf{b}, XRD pattern in the $(H\,K\,1.5)$ plane with simulated diffuse scattering. Data at $696\,$K in panels \textbf{a},\textbf{b} are from a different sample, scaled to the reference Bragg reflection at $725\,$K.
    \textbf{c}, Configuration of Ru-Ru dimers in the disordered phase, called A-type in the main text. In the average $Ibmm$ structure, A-type is aligned along the $b_\mathrm{o}$ axis [Fig.~\ref{fig:fig1}\textbf{d}].
    \textbf{d}, Another configuration of Ru-Ru dimers, referred to as B-type, with $\pi$-shifted displacements compared to A-type. Random arrangements of the two types along the $b_\mathrm{o}$ axis produce the calculated diffraction patterns in panels \textbf{a},\textbf{b}.
    }
    \label{fig:fig3}
  \end{center}
\end{figure*}
\begin{figure}[t]
  \begin{center}
		\includegraphics[clip, trim=0cm 0cm 0cm 0cm, width=1.\linewidth]{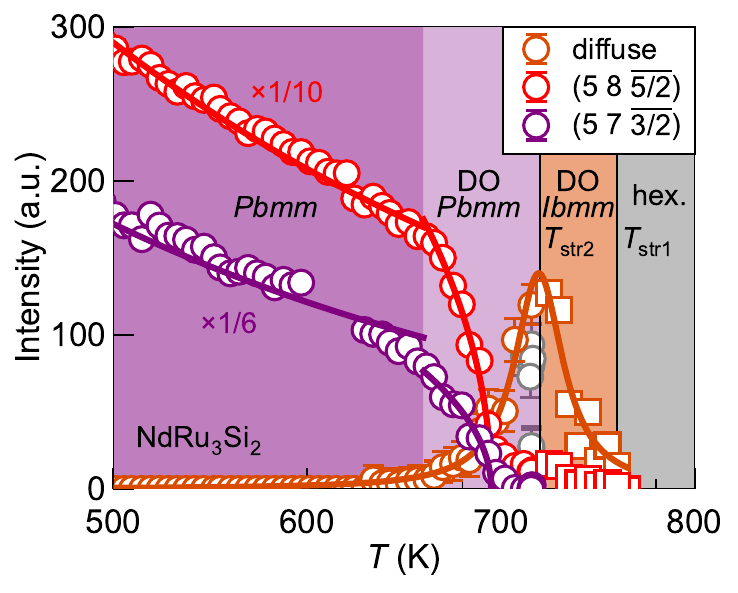}
    \caption[]{(color online). Critical behavior and effects of Debye-Waller factor in single-crystal X-ray diffraction intensities of \NRS{}. Gray, orange, and purple regions express hexagonal (hex.) $P6/mmm$, disordered (DO) $Ibmm$, and $Pbmm$ phases. The thin purple area indicates the disordered $Pbmm$ phase. Orange, red, and purple open circles represent diffuse scattering, $(5\ 8\ \overline{5/2})$ reflection, and $(5\ 7\ \overline{3/2})$ reflections in an orthorhombic basis without $c$ axis doubling, respectively. The diffuse scattering is taken at $(1/4\, 9/4\, \overline{5/2})$ where no Bragg reflection is expected. The orange line is a guide to the eye. The red and purple lines are fits to the Bragg intensities with a Debye-Waller factor and scaling functions. Open squares are from a different dataset, as explained in Methods. Clear outliers are colored in gray.
    }
    \label{fig:fig4}
  \end{center}
\end{figure}
\begin{figure*}[t]
  \begin{center}
		\includegraphics[clip, trim=0cm 0cm 0cm 0cm, width=1.\linewidth]{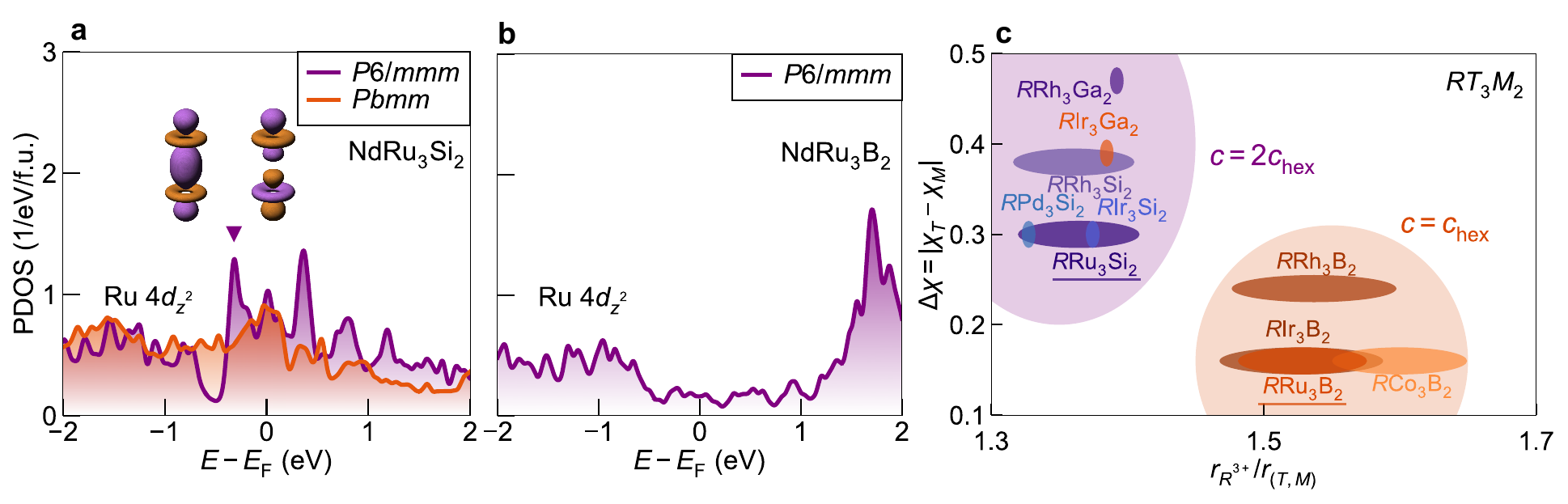}
    \caption[]{(color online). Molecular orbital formation of kagome $d_{z^2}$ orbitals and generalization to $RT_3M_2$ kagome metals ($T$ = transition metal, $M$ = main-group element).
    \textbf{a},\textbf{b}, Partial density of states (POS) of Ru-4$d_{z^2}$ orbitals for \NRS{} and \NRB{}, calculated by density functional theory. Purple and orange colors represent $P6/mmm$ and $Pbmm$ phases, respectively. The purple triangle indicates the PDOS peak, which is removed by molecular orbital formation between kagome layers. Schematics of bonding and anti-bonding states are shown as insets.
    \textbf{c}, Classification of $RT_3M_2$ kagome metals based on electronegativity difference and atomic radii. $\chi_T\ (\chi_M)$ is electronegativity of $T$ ($M$) and $\Delta \chi=|\chi_T - \chi_M|$ is the difference between them. $r_{R^{3+}}$ and $r_{(T,M)}$ are the ionic radius of $R^{3+}$ and the average atomic radius of $T$ and $M$. The tolerance factor is defined by their ratio. The orange region includes hexagonal and monoclinic structures with $c$ equal to that of the pristine hexagonal phase ($c_\mathrm{hex}$), while the purple area contains orthorhombic structures with $c = 2c_\mathrm{hex}$.
    }
    \label{fig:fig5}
  \end{center}
\end{figure*}

\section{Introduction}
The unique geometry of the kagome lattice, a network of corner-sharing triangles, gives rise to characteristic electronic structures, namely, Dirac cones, flat bands (FB), and van Hove singularities (vHs)~\cite{Yin2022-ps,Checkelsky2024-ik}. Our materials of interest, \RRS{} and \RRB{}, where Ru atoms form the kagome lattice [Fig.~\ref{fig:fig1}\textbf{a}], host FB and vHs near the Fermi level as shown in Fig.~\ref{fig:fig1}\textbf{b},\textbf{c}, respectively. Kagome metals can generally be categorized into four primary structural families, each distinguished by its specific stacking arrangement of triangular, honeycomb, and kagome lattices~\cite{Jiang2025-rh,Junze2025-pd}. These families include the \FG{}-type ($1$-$1$)~\cite{Teng2022-mu}, \KVS{}-type ($1$-$3$-$5$)~\cite{Ortiz2019-pr}, \SVS{}-type ($1$-$6$-$6$)~\cite{Arachchige2022-ai}, and \LRS{}-type ($1$-$3$-$2$)~\cite{Barz1980-ep,Plokhikh2024-eg}. Each family exhibits distinct structural motifs, which, in turn, influence their electronic landscapes and resulting physical properties. Among these kagome metals, \LRS{} stands out as a unique flat-band host exhibiting the highest superconducting transition temperature of $T_\mathrm{c}=7.8\,$K~\cite{Barz1980-ep,Mielke2021-ha,Junze2025-pd}. The superconducting state is proposed to be driven by mode-selective coupling between kagome $d_{x^2-y^2}$ orbitals and kagome phonons~\cite{Junze2025-pd,Misawa2025-xv}.

In these broad classes of kagome metals, orthorhombic distortions are common and are often either suppressive of or accompanied by charge density wave (CDW) order—an emerging area of research.
In fact, CDWs in kagome materials are of significant interest: a prominent example is the $3\bm{q}$ CDW order on the kagome lattice observed in \AVS{} (\textit{A}=K, Rb, Cs)~\cite{Wilson2024-ca} with potentially rotational symmetry and time-reversal symmetry breaking order well below the CDW transition~\cite{Zheng2022-ly,Guo2022-me,Guo2024-ra}. The electronic instability associated with vHs is considered a primary mechanism for the formation of these highly symmetric charge orders~\cite{Kiesel2013-kr,Wang2013-ha,Park2021-eo,Denner2021-ye,Tazai2022-qw,Tazai2023-sn}. On the other hand, the $\sqrt{3} \times \sqrt{3}$-type CDW order observed in \SVS{} is among the most commonly encountered examples~\cite{Arachchige2022-ai,Cao2023-qg,Korshunov2023-fg}. Soft phonons are present in the $k_z=\pi$ plane, and this instability is explained by the out-of-plane displacements of one-dimensional chains between a transition metal ($T$) and a main-group element ($M$)~\cite{Korshunov2023-fg,Feng2024-hz}. The $\sqrt{3} \times \sqrt{3}$ order is also reported in LuNb$_6$Sn$_6$ and \textit{T}Co$_3$B$_2$ (\textit{T}=Zr, Hf)~\cite{Ortiz2025-ca,Voroshilov1971-no}.

On the other hand, many kagome metals distort, often above room temperature, into orthorhombic or monoclinic structures with low symmetry. In the $1$-$6$-$6$ family, versatile low-symmetry structures are reported at room-temperature~\cite{Venturini2006-ma,Venturini2008-ov}. The orthorhombic distortion is attributed to displacements of $M-T$ chains along the $c$ axis through phonon calculations~\cite{Feng2024-hz}. As for the $1$-$1$ family, FeGe undergoes a monoclinic distortion below the magnetic transition, and subsequently increases its symmetry to orthorhombic upon entering the CDW phase~\cite{Wu2024-ck}. In the $1$-$3$-$2$ family, all the known silicides exhibit orthorhombic structures at room temperature~\cite{Cenzual1988-is,Verniere1995-ik,Shigetoh2007-tx,Sharma2023-xp}. As compared to the undistorted $P6/mmm$ structure in Fig.~\ref{fig:fig1}\textbf{a}, kagome metals with orthorhombic distortion have a two-times expanded $c$ axis lattice constant. Despite the formation of a superstructure, this is a typical structural phase transition without Fermi surface nesting, rather than a genuine CDW order, as discussed in this study. A recent example is \LRS{} which further has two CDW transitions in the orthorhombic phase~\cite{Plokhikh2024-eg,Mielke2024-as}. One of these is similar to the one in the recently discovered CsCr$_3$Sb$_5$, which exhibits superconductivity under pressure~\cite{Liu2024-cx}. So far, however, the crystal structures of the other \RRS{} remain unknown. While \textit{ab initio} calculations suggest soft phonon modes in the pristine hexagonal phase of \LRS{}~\cite{Plokhikh2024-eg,Junze2025-pd}, the underlying origin of this structural instability remains unresolved—not only in this material, but across broader classes of kagome metals.

In this study, we choose \RRS{} and \RRB{} as prototypical kagome metals with kagome-derived bands close to the Fermi energy. We investigate structural stability by synchrotron single-crystal X-ray diffraction (SXRD).
In \RRS{} (\textit{R}=Nd, Pr), we observe the structural phase transition from the parent $P6/mmm$ structure to the orthorhombic space group $Pbmm$ [Fig.~\ref{fig:fig1}\textbf{e}] via a critical regime with diffuse scattering [Fig.~\ref{fig:fig1}\textbf{d}]. The low-temperature structure is characterized by $c$ axis doubling, similar to other orthorhombic kagome metals~\cite{Chevalier1981-ln,Cenzual1988-is,Verniere1995-ik,Shigetoh2007-tx,Sharma2023-xp}. We observe Ru-Ru dimerization along this axis and thus ascribe the diffuse scattering in the intermediate-temperature range to disorder of the kagome dimers [Fig.~\ref{fig:fig1}\textbf{f}-\textbf{h}]. The disordered dimer model successfully captures the essential features of the diffuse scattering. Supported by electronic structure calculations, we attribute the origin of this phase transition to molecular orbital formation between the kagome $d_{z^2}$ orbitals.
By comparison with \RRB{} (\textit{R}=Pr, Gd, Lu), which preserve the ideal kagome lattice down to low temperatures, we propose a framework for assessing the structural chemistry of kagome metals. The present framework, based on electronegativity and atomic radii, may be applicable to a wide variety of kagome metals beyond the $1$-$3$-$2$ systems. Our findings offer a guiding principle for the design of tailored kagome compounds with hexagonal symmetry.

\section{Results}
\subsection{Sequence of orthorhombic distortions in \RRS}
We perform SXRD on \NRS{} and \PRS{} to reveal their structural properties. Since they behave similarly, we focus on \NRS{} in the main text.
Above $T_\mathrm{str1} = 760\,$K, \NRS{} exhibits an XRD pattern indicating the hexagonal structure with $P6/mmm$ symmetry [Fig.~\ref{fig:fig3}\textbf{a}].
There are no reflections in the $L=\,$half-integer planes, as shown in Fig.~\ref{fig:fig3}\textbf{b}.
Upon cooling below $T_\mathrm{str2} = 660\,$K, we observe a new set of Bragg reflections in both $L=$integer and half-integer planes, indicating a doubling of the $c$ axis.
Illustrated in Fig.~\ref{fig:fig3}\textbf{a},\textbf{b} are the XRD patterns taken at $30\,$K, showing that the new reflections are indexed by the orthorhombic $Pbmm$ structure with three $C_3$ symmetric domains. From the structural refinement, we estimate that the volume fraction of orthorhombic $Pbmm$ domains is $0.0336(6),\ 0.6547(4),\ 0.3116(4)$, respectively~\cite{SI}. We note that the resolution of the 2D detector used in this study does not allow us to observe the splitting of Bragg reflections associated with the orthorhombic distortion. However, the XRD pattern and intensity distribution of the satellite reflections confirm the breaking of six-fold symmetry, similar to \LRS{}~\cite{Plokhikh2024-eg}. Consistent with our structure refinement, optical birefringence measurement visualizes a large orthorhombic domain on a single crystal of \NRS{}~\cite{SI}.

We also perform SXRD on \RRB{} (\textit{R}=Pr, Gd, and Lu) to compare the structural phase transition with \RRS{}.
These materials do not show any structural transition from $P6/mmm$ down to $30\,$K~\cite{SI}.

\subsection{One-dimensional disorder of kagome dimers}
In the intermediate regime between $T_\mathrm{str1}$ and $T_\mathrm{str2} = 720\,$K, we observe diffuse scattering in the $L=\,$half-integer planes as well as orthorhombic Bragg reflections. First, ignoring the weak diffuse scattering and focusing on the Bragg reflections, the space group is determined as $Ibmm$~\cite{SI}; see Fig.~\ref{fig:fig3}\textbf{a},\textbf{b}. The refined domain ratio is $0.023(15)$, $0.534(10)$, $0.443(10)$, comparable to the $Pbmm$ phase.
        
We then analyze the diffuse scattering, which is subsequently attributed to the formation of Ru-Ru dimers observed in the low-temperature phase. Key features are (i) its presence along the $b^*_\mathrm{o}$ axis with $C_3$ symmetric directions arising from the other domains, and (ii) its absence in the $L=\,$integer or $H=\,$even planes. On one hand, (i) suggests that the structure is long-range correlated in the $a_\mathrm{o}c_\mathrm{o}$ plane but not along the $b_\mathrm{o}$ axis in real space. On the other hand, (ii) indicates that two sublattices connected by a $1/2$-translation along the $c_\mathrm{o}$ axis or $a_\mathrm{o}$ axis must have opposite displacements.
From the refinement of the average structure in Fig.~\ref{fig:fig1}\textbf{d},\textbf{g}, the conditions on sublattices are satisfied by two Ru atoms, Ru$_1$ and Ru$_2$, which dimerize between adjacent kagome layers. To account for the diffuse scattering, we thus build a disordered dimer model on the kagome Ru sites based on the $Ibmm$ structure. In the average $Ibmm$ structure, the dimers labeled as A-type in Fig.~\ref{fig:fig3}\textbf{c} are aligned along the $b_\mathrm{o}$-axis. The diffuse scattering thus arises from the random arrangement of A-type and $\pi$-shifted B-type in this direction [Fig.~\ref{fig:fig3}\textbf{d}]. The distortion amplitude has to be the same for both types to account for the absence of diffuse scattering in $L=$integer planes. The intensity of both the diffuse scattering and Bragg peaks in the $L =$half-integer planes is governed by the distortion amplitude and the volume fractions of the two configurations, $v_\mathrm{A}$ and $v_\mathrm{B}$. When $v_\mathrm{A} = v_\mathrm{B} = 0.5$, the distortions cancel out on average, and no Bragg reflections are expected in the $L=$half-integer planes. Therefore, we set $v_\mathrm{A} = 0.7$ and randomly distribute the two configurations along the $b_\mathrm{o}$ axis. We then simulate the diffraction patterns in Fig.~\ref{fig:fig3}\textbf{a},\textbf{b}, which agree well with the observations for \NRS{}.
\subsection{Long-range orthorhombic phase stabilized by dimer recombination}
Upon cooling, the one-dimensional disorder is suppressed through a recombination of the dimers and a concomitant reduction in entropy. Below $T_\mathrm{str2}=720\,$K, new diffraction spots, consistent with the $Pbmm$ space group, gradually emerge. Nevertheless, the diffuse scattering persists until around $660\,$K, indicating the continued presence of disorder. The eventual transition to the long-ranged $Pbmm$ structure is driven by a reorganization of dimer bonds: one of the Ru$_1$-Ru$_1$ or Ru$_2$-Ru$_2$ dimers is replaced by new dimers involving the third, central Ru atom (Ru$_3$) [Fig.~\ref{fig:fig1}\textbf{h}]. Importantly, this reorganization is enabled by the presence of three Ru sublattices within each kagome unit, a feature unique to the kagome lattice. Such triadic connectivity allows for a rich dimer landscape and plays a key role in the transition mechanism.

We note that Ru$_3$ atoms are not disordered in the intermediate temperature range where the diffuse scattering is active. Suppose the disorder along the $b_\mathrm{o}$ axis, Ru$_3$ atoms at $(x=1/2,\, y=0)$ and $(x=0,\, y=1/2)$ [Fig.~\ref{fig:fig1}\textbf{d}] must be displaced in the parallel or antiparallel direction with finite probability, similar to other Ru atoms.
Therefore, the diffuse scattering should appear both in $H=\,$even and odd planes, which violates the condition (ii). For the same reason, Nd atoms do not fluctuate either, because the in-plane geometry of Nd is essentially the same as that of Ru$_3$ atoms (triangle), differing only by a lattice translation [Fig.~\ref{fig:fig1}\textbf{d},\textbf{e}].
\subsection{Critical behavior of XRD intensities}
To track the successive structural transitions, we perform temperature-dependent SXRD.
In Fig.~\ref{fig:fig4}, we show the diffuse scattering intensity in blue and the Bragg reflections characterizing the $Ibmm$ and $Pbmm$ structures in red and purple, respectively.
The diffuse scattering starts to evolve at $T_\mathrm{str1}=760\,$K, at the same time as the $Ibmm$-type Bragg reflections, and takes a maximum at $T_\mathrm{str2}=720\,$K, from which it decreases and completely disappears at around $660\,$K.
Cooling below $T_\mathrm{str2}=720\,$K, the diffuse scattering is weakened while the intensity of $Pbmm$ Bragg reflections gradually increases: Dimerization of the central Ru atom in each triangle breaks the body-centring symmetry of $Ibmm$.

The temperature dependence of the Bragg intensities well below the critical temperatures is entirely attributed to the Debye-Waller factor, $I=I_0\exp(-2W)$, with $W = 3\hbar^2|\bm{G}|^2T/\left(2Mk_\mathrm{B}\Theta_\mathrm{D}^2\right)$ where $\bm{G}$, $M$, $\Theta_\mathrm{D}$ are the reciprocal lattice vector, the mass of an atom (Ru), and the Debye temperature. From this equation and a fit to the temperature dependent intensity of ($5\ 8\ \overline{5/2}$) and ($5\ 7\ \overline{3/2}$), we estimate $\Theta_\mathrm{D} = 154\,$K and $182\,$K, respectively, which are comparable to $\Theta_\mathrm{D} = 194\,$K from the analysis of the low-temperature specific heat in \LRS{} when including a correction for electronic correlations~\cite{Li2011-qv}.

In the intermediate temperature range, critical fluctuations play a role. For the diffuse scattering, we observe a diverging behavior of the intensity $I(T)$ towards $T_\mathrm{str2}$. The diffuse scattering is related to susceptibility, and thus this critical behavior follows the power law $I\propto |t|^{-\gamma}$, where $t$ is the reduced temperature defined as $t = (T-T_\mathrm{str2})/T_\mathrm{ste2}$ and $\gamma$ is the critical exponent of the susceptibility~\cite{Trenkler1998-kx}. Fitting to this functional form, we obtain $\gamma=1.1(2)$, consistent with the mean-field value ($\gamma=1.0$). As for the Bragg reflections, their intensities gradually develop and then follow the scaling behavior of an order parameter with the mean-field critical exponent $\beta=1/2$. Our differential scanning calorimetry measurements confirm an anomaly at $722\,$K~\cite{SI}, consistent with $T_\mathrm{str2}$ in SXRD.
\subsection{Molecular orbital formation as the origin of successive orthorhombic distortions}
So far, we have revealed the successive orthorhombic distortions in \RRS{} driven by the consecutive dimerization of kagome Ru atoms and the absence of such a transition in \RRB{}.
The dimers along the $c_\mathrm{o}$ axis in \RRS{} imply molecular orbital formation of the kagome $d_{z^2}$ orbitals. To validate this scenario, we perform \textit{ab initio} calculations of the partial density of states (PDOS) of Ru-$d_{z^2}$ for \NRS{} and \NRB{}.
We find that the PDOS peak in the $P6/mmm$ phase of \NRS{}, located near the Fermi level, is removed after the structural transition to the $Pbmm$ phase, as illustrated in Fig.~\ref{fig:fig5}\textbf{a}.
This is consistent with the molecular orbital formation of kagome $d_{z^2}$ orbitals. A competing scenario for the high-temperature phase transition is Fermi surface nesting with a related electronic instability. This is excluded from first-principle calculations for the Fermi surface of \NRS{}~\cite{SI}.
On the other hand, almost no $4d_{z^2}$ PDOS is observed for \NRB{} around the Fermi energy [Fig.~\ref{fig:fig5}\textbf{b}]. These features are also evident in the orbital projected band structure of \NRS{} and \NRB{}, depicted in Fig.~\ref{fig:fig1}\textbf{b} and \textbf{c}, respectively.

We attribute the difference in the Ru-$d_{z^2}$ orbitals' contribution to the variation in covalent bonding strength between Ru and the surrounding B or Si atoms. Compared to silicon’s electronegativity ($\chi_\mathrm{Si} = 1.9$), boron’s ($\chi_\mathrm{B} = 2.0$) is closer to that of ruthenium ($\chi_\mathrm{Ru} = 2.2$), and thus Ru-$d_{z^2}$ orbitals form stronger covalent bonds with unoccupied B-$p$ orbitals. This bonding interaction pushes the Ru-$d_{z^2}$ states away from the Fermi level in \RRB{}. In contrast, in the silicon analog, the weaker bonding between Ru-$d_{z^2}$ and Si-$p$ orbitals allows the $d_{z^2}$ states to remain near the Fermi level, making them available for dimer formation. Comparison of the PDOS for Ru-$d_{z^2}$ and B-/Si-$p_z$ orbitals supports this scenario~\cite{SI}.
\subsection{Molecular orbital formation in $1$-$3$-$2$ kagome metals}
Generalizing to the entire $1$-$3$-$2$ family of kagome metals, hexagonal structure are favored when (i) the difference in electronegativity $\Delta \chi$ between the transition metal ($T$) and main-group element ($M$), defined as $\Delta \chi = |\chi_T - \chi_M|$, is small and (ii) the tolerance factor, defined as the ratio of the ionic radius of $R^{3+}$ and the average radius of $T$ and $M$ atoms, expressed as $t = r_R^{3+}/r_{(T, M)}$, is moderately large. For the $1$-$3$-$2$ family, we have $r_{(T, M)} = (3r_{T} + 2r_{M})/5$. We thus plot these two parameters for various $1$-$3$-$2$ kagome metals in Fig.~\ref{fig:fig5}\textbf{c} ~\cite{Niihara1973-vo,Ku1980-nl,Chevalier1981-ln,Cenzual1988-is,Verniere1995-ik,Sologub2003-vd,Dubman2005-ap,Shigetoh2007-tx,Seidel2015-rc,Manni2019-ji,Gui2022-mq,Sharma2023-xp}. The plot demonstrates that the borides have lower $\Delta \chi$ and larger $t$ and thus commonly exhibit the absence of $c$ axis doubling or orthorhombic distortions, while the opposite is true for all silicides and gallides. In the $AB_5$-type ($1$–$5$) family, which is isostructural to the $1$–$3$–$2$ family upon replacing the $M$ site with $T$ atoms, the stability range has been shown to lie within $1.30 \leq t = r_A / r_B \leq 1.77$~\cite{Raynor1977-lh}. Figure~\ref{fig:fig5}\textbf{c} confirms the same behavior in the $1$-$3$-$2$ systems. As silicides and gallides are in close proximity to the lower limit of $t$, a distorted structure may be preferred. The gallide LaIr$_3$Ga$_2$ stands out as an exceptional hexagonal system, likely due to its strong two-dimensionality: $c/a = 0.7$.

Focusing on \RRS{}, the crystal symmetry decreases and the structure becomes less stable as the rare-earth atomic number increases. While \LRS{} forms a base-centered $Cccm$ structure, the heavier \RRS{} compounds (Pr, Nd) adopt a primitive $Pbmm$ structure. The trend is further supported by our differential scanning calorimetry measurements, which show that $T_\mathrm{str2}$ increases with heavier rare-earth elements in \RRS{}~\cite{SI}. This is because heavier rare-earth elements have smaller ionic radii, which push the system closer to the lower limit of the tolerance factor. On the other hand, the other silicides stabilize in the body-centered $Ibmm$ structure. The body-centering symmetry is broken by the recombination of kagome dimers [Fig.~\ref{fig:fig1}\textbf{g},\textbf{h}]. Whether this happens likely depends on subtle details of crystal parameters, which control the competition between energy loss from lattice deformation and energy gain from molecular orbital formation.

We note that a previous study evaluated the effects of the electronegativity difference between $A$ and $B$ in the $1$-$5$ systems, concluding that it does not have a major impact on phase stability~\cite{Raynor1977-lh}. However, its impact on ortho\textbf{}rhombic distortions was not discussed. Indeed, our focus is on the electronegativity difference between the $T$ and $M$ sites, which are, in contrast, occupied by the same element in the $1$-$5$ family.

\section{Conclusion}
In summary, we observe a sequence of orthorhombic structural phase transitions in \RRS{} and its absence in \RRB{}, utilizing synchrotron SXRD.
Through modeling of diffuse scattering, precise structural refinement, and \textit{ab initio} calculations, we attribute the distortions to molecular orbital formation of kagome $d_{z^2}$ orbitals.
This behavior appears to be a common feature of $1$-$3$-$2$ kagome metals, as we find that the orthorhombic distortion is correlated with the electronegativity difference between a transition metal and a main-group element and the tolerance factor defined as the atomic radius ratio.

Although similar diffuse scattering was reported in \LRS{}, it was not modeled and not attributed to Ru dimers~\cite{Plokhikh2024-eg}. Contrary to \NRS{}, \LRS{} does not exhibit Bragg reflections, but shows diffuse scattering at $L=$half-integer planes in the critical regime between the $P6/mmm$ and $Cccm$ phases. Our model explains this behavior in the case where $v_\mathrm{A} = v_\mathrm{B}$; only diffuse scattering is present because distortions are canceled out on average. The equal occupation of the two configurations is consistent with the low-temperature \textit{Cccm} phase, which exhibits a long-range alternation of A- and B-type units within the unit cell. Furthermore, as a precursor of charge-density wave transitions, other types of diffuse scattering were reported in FeGe and ScV$_6$Sn$_6$~\cite{Cao2023-qg,Korshunov2023-fg,Subires2025-dh}. However, their origin is different from \RRS{}: for the former, it is Ge atoms on the honeycomb lattice that dimerize, while for the latter, it is likely a one-dimensional $T-M$ chain -- although it was not modeled. 
Our disordered dimer model thus stands out by describing a collective instability inherent to the kagome lattice, where the presence of three distinct sublattices gives rise to a diverse landscape of competing dimer configurations. We note that similar orthorhombic structures are common in other families of kagome metals, such as $1$-$3$-$5$, $1$-$1$, and $1$-$6$-$6$ families~\cite{Venturini2006-ma,Venturini2008-ov,Liu2024-cx,Feng2024-hz,Wu2024-ck}, underlining the potential applicability of our framework to these material classes, and its utility for designing kagome metals with high rotational symmetry.

\textbf{Note added:} After the submission of this work, we became aware of a preprint reporting the crystal structure of \YRS{} as a charge-ordered phase.~\cite{Kral2025-xs}. The space group is the same as \RRS{} (\textit{R}=Pr, Nd), characterized by interlayer dimerization of kagome atoms, and we attribute the origin of orthorhombic distortions to molecular orbital formation.

\section{Methods}
\subsection{Crystal growth}
\RRS{} and \RRB{} are synthesized by the arc melting technique in a high-purity argon atmosphere.
For \RRS{}, to avoid the competing magnetic impurity \textit{R}Ru$_2$Si$_2$, we add excess Ru to the starting materials~\cite{Barz1980-ep}. Increasing excess Ru is required to obtain \RRS{} with heavier rare-earth elements. \RRB{} is melted from a stoichiometric combination of elements. Small single crystals, approximately $50\,\mu\mathrm{m}$ in size and suitable for SXRD measurements, are obtained by mechanical fragmentation of the crystals. A $1\,$mm$^3$ single crystal of \NRS{}, with around $15$\% of Ru impurity, for optical measurements is found from an ingot after the Czochralski growth.

\subsection{Single-crystal X-ray diffraction}
SXRD is performed at BL$02$B$1$ of the synchrotron X-ray facility, SPring-$8$ (Japan).
Diffraction patterns are recorded using a CdTe PILATUS area detector.
Integrated intensities are collected in the CrysAlisPro program~\cite{Agilent-Technologies-Ltd2014-hl}. Equivalent reflections are averaged, and structure parameters are refined in Jana$2006$~\cite{Petricek2014-pc}. In the space group descriptions (e.g., \(Ibmm\)), the axes are retained as in the pristine structure and are not transformed to the conventional setting (e.g., \(Imma\)) to facilitate direct comparison.
In Fig.~\ref{fig:fig4}, the open circles represent data from one fixed frame during a temperature scan, while the squares correspond to intensities from the reconstructed reciprocal space in a dataset with an extended momentum scan. The latter are normalized using the intensity of a Bragg reflection common to both datasets. Crystal structures are drawn by VESTA~\cite{Momma2011-gm}.

\subsection{Modeling of diffuse scattering}
Diffuse scattering is modeled by considering the disorder of Ru atoms in the $Ibmm$ phase, as illustrated in Fig.~\ref{fig:fig3}\textbf{c},\textbf{d}. We arrange A- and B-type with the probability $v_\mathrm{A}$ and $v_\mathrm{B}$ along the $b_\mathrm{o}$ axis and constructs a supercell comprising \(L_x \times L_y \times L_z\)=\(60 \times 60 \times 1\) orthorhombic unit cells, each containing $12$ Ru atoms. We set 
$v_\mathrm{A} = 0.7$ in the main text. To mitigate finite-size effects, the mesh spacing is set to $2\pi / a_\mathrm{o}L_x\,$(\AA) and $2\pi / b_\mathrm{o}L_y\,$(\AA) along the $k_x$ and $k_y$ directions, respectively. This results in a $679 \times 1173$ mesh for computing reciprocal space maps at $L =1.5$ and $3$. To suppress unphysical coherence across the entire crystal, we calculate intensities over a randomly sampled volume of $10 \times 10 \times 1$ unit cells and average the results over 100 samples. The total diffraction pattern is obtained by summing the intensities from three orthorhombic domains. For simplicity, equal occupation of domains is assumed. The atomic form factor of Ru is included in the calculation using the Cromer–Mann Gaussian parametrization. To enable direct comparison, the calculated diffraction patterns are linearly rescaled to the experimental mesh, as shown in Fig.~\ref{fig:fig3}\textbf{a},\textbf{b}. Contributions from the other elements are not found to be necessary to reproduce the experimentally observed diffuse scattering.

\subsection{Specific heat measurement}
Differential scanning calorimetry (DSC) is performed over the temperature range of $300\,$K to $850\,$K using a simultaneous thermal analyzer (STA $449$ F$1$ TG-DSC, Netzsch) under a nitrogen atmosphere. DSC signals are then converted to specific heat~\cite{SI}.

\subsection{First-principles calculations}
First-principles calculations are performed using the Vienna Ab initio Simulation Package (VASP) $6$ with the Perdew-Burke-Ernzerhof (PBE) generalized gradient approximation for the exchange-correlation functional~\cite{Perdew1996-ih}. The projector augmented-wave (PAW) pseudopotentials employed correspond to Nd, $\mathrm{Ru}_\mathrm{pv}$, Si, and B~\cite{Blochl1994-zh,Kresse1999-ww}. A $\Gamma$-centered $15 \times 15 \times 19$ Monkhorst--Pack grid is used for sampling the Brillouin zone. Spin-orbit coupling, as implemented in VASP, is included in all calculations~\cite{Steiner2016-ts}. Structural relaxations are performed using the residual minimization method with direct inversion in the iterative subspace (RMM-DIIS), employing a step size of $0.25$~\cite{Pulay1980-za}.

\subsection{Optical birefringence measurement}
Optical birefringence is measured for real-space observation of orthorhombic domains using an optical microscope with polarizer and analyzer plates~\cite{Xu2022-vi}. An LED light source (M$970$L$4$, Thorlabs) with a center wavelength of $970\,$nm and a bandwidth of approximately $50\,$nm is used. Linearly polarized, near-infrared collimated light is directed onto the sample whose surface is perpendicular to the $c$ axis. The reflected light is then collected and passed through an analyzer before reaching a CMOS camera (CS$505$MO, Thorlabs), which captures a polarization-resolved image of the sample. The incident polarization angle dependence of the polarization rotation angle is measured by rotating the sample while fixing the incident polarization.

\subsection{Acknowledgement}
This work was supported by JSPS KAKENHI Grants No. JP22K20348, No. JP23H05431, No. JP23K13057, No. JP24H01607, and No. JP24H01604, as well as JST CREST Grant Nos. JPMJCR1874 and JPMJCR20T1 (Japan) and JST FOREST Grant No. JPMJFR2238 (Japan). It was also supported by Japan Science and Technology Agency (JST) as part of Adopting Sustainable Partnerships for Innovative Research Ecosystem (ASPIRE), Grant Number JPMJAP2426. Work at Princeton was supported by the David and Lucille Packard foundation, the Princeton Catalysis Initiative (PCI), and by the Air Force office of Scientific Research under award number FA9550-24-1–011. The synchrotron single-crystal X-ray experiments were performed at BL02B1 in SPring-8 with the approval of RIKEN (Proposal No. 2024A1760, 2024B1839, and 2024B2010). The in-house single-crystal X-ray experiments were supported by ``Advanced Research Infrastructure for Materials and Nanotechnology in Japan (ARIM)'' of the Ministry of Education, Culture, Sports, Science and Technology (MEXT), Grant Number 24UT0147.
M.J. acknowledges the computing resources provided by North Carolina State University High Performance Computing Services Core Facility (RRID:SCR{\_}022168).
\bibliography{ref}
\end{document}